# Towards Disruption Tolerant ICN

Hasan M A Islam, Andrey Lukyanenko, Sasu Tarkoma, and Antti Yla-Jaaski
Aalto University
Espoo, Finland
Email: firstname.lastname@aalto.fi

*Abstract*—Information-Centric Networking (ICN) is a prominent topic in current networking research. ICN design significantly considers the increased demand of scalable and efficient content distribution for Future Internet. However, intermittently connected mobile environments or disruptive networks present a significant challenge to ICN deployment. In this context, delay tolerant networking (DTN) architecture is an initiative that effectively deals with network disruptions. Among all ICN proposals, Content Centric Networking (CCN) is gaining more and more interest for its architectural design, but still has the limitation in highly disruptive environment. In this paper, we design a protocol stack referred as CCNDTN which integrates DTN architecture in the native CCN to deal with network disruption. We also present the implementation details of the proposed CCNDTN. We extend CCN routing strategies by integrating Bundle protocol of DTN architecture. The integration of CCN and DTN enriches the connectivity options of CCN architecture in fragmented networks. Furthermore, CCNDTN can be beneficial through the simultaneous use of all available connectivities and opportunistic networking of DTN for the dissemination of larger data items. This paper also highlights the potential use cases of CCNDTN architecture and crucial questions about integrating CCN and DTN.

*Keywords—Content Centric Networking (CCN), Delay Tolerant Networking (DTN).*

## I. INTRODUCTION

Over the past few years, the vast majority of Internet usage is dominated by content distribution and retrieval with a large amount of digital content. Whereas, the conventional Internet architecture is still based on end-to-end communication model and thus faces a number of recognized limitations, for instance, decoupling address from an end-point identity, mobility and disruption tolerance, and above all, scalable and efficient content distribution. In response to these, Information-Centric Networking (ICN) (e.g., CCN [1], DONA [2], PURSUIT/PSIRP [3], [4], [5], NetInf [6]) emerges a paradigm shift from host centric to information centric communication model, i.e., *Named data* is the central element of ICN communication, instead of its physical location. Furthermore, all ICN representatives share the following key features:

- In-network caching, i.e., the data is cached within the transport network.
- Content based addressing, i.e., addressing scheme is based on content names rather than pointing to a specific location.
- Content based security, i.e., securing the content itself rather than communication channel.
- Receiver-driven communication, i.e., consumer requests for a content that is routed to the nearest copy of the content. Subsequently, if available, a node on the forwarding path replied the content back to the originating requester.

Numerous articles [7], [8], [9] delineate arguments for ICN, whether such changes are required in internetworking architecture or not, and pose several research agenda. However, as presently Internet traffic is mostly data oriented (e.g., google, facebook, amazon, ebay), the design principles of ICN seem more beneficial and provide useful features to end users. For instance, the end users simply need to express their interest for a particular data, instead of having a reference to specific location where data is to be retrieved from.

Nevertheless, intermittently connected network topology or network disruption incurs a significant challenge for ICN deployment. For instance, name resolution may fail due to network disruptions especially when the elements of distributed resolution services are affected by network partitioning. Moreover, name resolution may become unreliable particularly when names are resolved to locator(s) that do not exist anymore. In fact, ICN development is in its early stage. ICN, however, should consider network disruptions in their designs.

In contrast, Delay-tolerant networking (DTN) [10], [11] is an initiative that introduces an architecture for challenged environment, which is particularly characterized by long delay paths, frequent unpredictable disconnections and network partitions. This architecture provides a flexible and resilient protocol for such networks. DTN is based on store-and-forward model utilizing persistent storage that is well distributed throughout the network. All data are cached in the network until an opportunistic contact is available to forward data. In particular, content based routing has been explored in DTN architecture [12]. Interestingly, DTN architecture assimilates some properties (e.g., in-network caching, late binding) of ICN designs or vice versa.

In this paper, we propose a protocol stack integrating ICN and DTN architecture to deal with network disruptions. For such, we first explore the differences and commonalities of ICN and DTN architectures. Next, we design and implement the protocol stack integrating CCN and DTN. This protocol stack enhances CCN routing strategy through Bundle Protocol of DTN architecture. Furthermore, we present our implementation experiences, and highlight the possible enhancements for future work.

The rest of this paper is organized as follows: Section II presents existing ICN approaches to disruption tolerance, alongside the limitations of these approaches in DTN. Section III describes our proposed architecture and implemen-tation details. Section IV describes content distribution and retrieval through our prototype in a mixed environment. Section V discuss the crucial questions about integrating CCN and DTN. Potential use cases are discussed in Section VI. Finally, we present related works in Section VII and conclude our paper in Section VIII.

## II. DISRUPTION TOLERANCE IN ICN

In general, ICN transports have three phases; content distribution, request for content, and respond data with matching request. Any of these phases in ICN transport can be interrupted by network disruptions. In this section, we briefly present the core aspects of ICN transports, alongside their disruption tolerance (e.g., large RTT, network partitioning).

### A. PSIRP/PURSUIT

In publish/subscribe (PSIRP/PURSUIT), contents are organized using scopes with an identification structure that provides hierarchically organized information [13]. This architecture decomposes the core network functions into three parts: scope specific *Rendezvous*, *Topology* and *Forwarding* functions (RTF functions). The Rendezvous function is responsible for matching subscription request with publications. The node where the requested content is found is called *Rendezvous point*. The Topology function monitors the network topology and creates information delivery path at different levels of inter-domain system. And the Forwarding function is responsible for forwarding contents. To publish a new content, publishers have to use two identifiers: a unique label for every piece of content referred as *rendezvous identifiers* RId and scope identifiers (SId). The publishers also need to locate the *rendezvous nodes* that are responsible for managing a scope and playing role of rendezvous point. Fundamentally, the forwarding is performed assigning a stack of path specific identifiers to the publication, referred as *Forwarding identifiers* (FId). FIds are determined for the links along the forwarding path. Each node along the path maintains a forwarding table that maps the incoming FId to its corresponding outgoing interface and identifier. However, this forwarding mechanism is not suitable for highly disrupted network, since the connections are sporadic. Furthermore, repeated unsubscribe and re-subscribe is not an optimal solution. The inclusion of rendezvous points is not adequate for delay tolerant environment.

### B. CCN

CCN communication is consumer driven, i.e., a consumer broadcasts interest packets over all available connectivities. Interest packets can also be forwarded via multicast or anycast based on the strategy of a CCN node. Each CCN node maintains three data structure; content store, pending interest table (PIT) and forwarding information base (FIB). Therefore, for each incoming interest packet on an interface, every node verifies its local cache for the matching content. Subsequently, if the node can satisfy the request, the content is sent back to the source(s) originating interests. Otherwise, the node forwards the interest packet to the interface(s) based on FIB table until the interest packet reaches a content source that can satisfy the interest. PIT keeps track of the ongoing interests so that the data can be sent back to the proper requester. CCN interests that are not satisfied within a reasonable amount of time are retransmitted. As CCN senders are stateless [1], the consumer is responsible for re-expressing interests if not satisfied. Intermediate node is responsible for retransmission on a particular interface. However, retransmission and re-expression of interests in highly long delay path may create redundant network traffic and consume a significant amount of bandwidth. Furthermore, PIT table may overflow with frequent disruptions and possible network collapse. In fact, this PIT bottleneck may raise inevitable constraints in terms of reliability and scalability. Thus, reverse path based on PIT and name aggregation are not suitable for the fragmented networks.

### C. DONA

DONA introduces a new class of entity called Resolution Handlers (RH), which are structured into a tree topology. In DONA, publishers publish data to local RH and only authorized publishers can register to RH prior to any publication of content objects. Content names are of P:L form, where P is the cryptographic hash of the publisher's public key and L is a label that identifies the content object. Each RH maintains a registration table to map a content name to next-hop RH towards the content objects. Requests for data are first routed to the consumer's local RH. If the local RH cannot satisfy the request, the request is forwarded up the tree topology until a source is found. However, data is sent back, if available, in response to requests either over reverse RH path (if RH provides universal cache infrastructure) or over a more direct route between the source and consumer (e.g., standard IP routing and Forwarding) [2]. Eventually, routing based on hierarchical RHs does not suit well with network disruptions, since routing table management for such environment needs frequent update to the registration table. Additionally, if standard IP routing and forwarding is utilized to exchange packets, DONA faces the same limitations of IP in such environment. In conclusion, as like CCN, DONA routing approaches pose a notable limitation in DTN.

### D. NetInf

NetInf transport uses both name-based routing and name resolution approaches. NetInf utilizes a *Convergence Layer* architecture to underpin communication on top of the underlying protocol in diverse scenarios. The inclusion of convergence layer architecture helps NetInf to access contents across the edge region incurring high disruptions (e.g., large RTT, sporadic connectivity). Moreover, name-based routing (Late Locator Construction (LLC)) in NetInf does not require state per networked object during the transportation. In contrast, routing based on name resolution services (NRS) (creating topology based on multi level DHT (MDHT)), in fact, will have the same limitations as like PSIRP/PURSUIT or DONA.

### E. DTN

In contrast, Delay tolerant network architecture significantly considers the limitations of traditional Internet architecture. Unlike IP, DTN deals effectively with long delays, high bit error rates, and network disruptions. In fact, DTN creates an opportunistic network on top of existing underlying Layer 2 and Layer 3 protocols. It weakens the necessity of an end-to-end path between source and destination for the duration of a communication session. For such, DTN utilizes store-and-forward mechanism over multiple paths and potentially long timescales to cope with highly disrupted environment. For instance, intermediate nodes store the messages until there is an opportunity to forward them to next hop. However, in contrast to ICN, DTN routing and forwarding still relies on end-point addressing. Table I summarizes the key features, differences and commonalties of ICN and DTN architectures.

| Criteria | CCN | NetINF | PSIRP | DONA | DTN |
|---|---|---|---|---|---|
| Naming | Hierarchical | Flat | Flat level identifiers within hierarchical scopes | Flat | general syntax of URI scheme |
| Name resolution | Route Aggregation via PIT | Multi-Level DHT/ Late Locator Construction(LLC) | Rendezvous service | Hierarchical RHs | Late binding |
| Routing | Name-based using PIT and FIB | NRS or name-based or Hybrid | using RTF function | Name-based (via RHs) | Next hop EID/ Lower layer addressing to Dest. |
| In Network caching | Yes | Yes | Limited to Rendezvous point | RH | bundle caching |
| Data forwarding | Route aggregation | Convergence layer (e.g., TCP, UDP) | Forwarding Function (e.g., Bloom Filter based forwarding | Existing Transport protocol | based on EID |

TABLE I: ICN and DTN features overview.

### III. PROTOTYPE DESIGN

In this section, we present the details of our prototype, specifically how to integrate CCN and DTN. We start by proposing a protocol stack for the native CCN and discuss the implementation details. Our prototype extends the native CCN transport by integrating the Bundle Protocol [14] of DTN architecture.

#### A. Motivation

The advancement and ubiquity of smartphones shift multimedia usage (e.g., video streaming) from personal computer to mobile phones. Content distribution and retrieval for mobile phones can be done by means of fixed infrastructure (e.g., cellular networks/access points). However, sending and receiving data can be interrupted by several reasons, such as network failure, battery limitations, network partitioning. Such scenarios fall into Pocket Switched Networking (PSN) that enables opportunistic network services for mobile users even in the absence of Internet connectivity. DTN architecture is very useful for such PSNs. Whereas, as discussed in Section II, all ICN tenants available today pose some limitations in network disruptions. These limitations of current ICN proposals in fragmented networks and the necessity of disruption tolerance persuade us to carry out research towards the interoperability of ICN and DTN.

#### B. Why CCN and DTN?

DTN and CCN share some commonalities in their designs (e.g., in-network caching, late binding). Conversely, there exist fundamental differences between CCN and DTN. For instance, DTN routing and forwarding is still relies on the conventional addressing scheme of senders and receivers.

On the other hand, CCN pose some limitations in disruptive networks. For instance, reverse path based on PIT in CCN fails particularly when the connections intermittently goes up and down. Moreover, next hop may not be available for a significant amount of time.

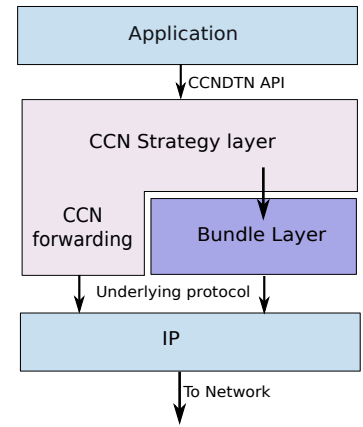

Fig. 1: CCNDTN Protocol Stack.

However, among all the ICN proposals, CCN is gaining more and more interest for its architectural design. Apart from IP, CCN introduces two new layers; *strategy* and *security* [1]. The strategy layer, in fact, provides flexibility to operate on top of IP or Layer 2 protocols. In addition, the strategy layer can utilize multiple simultaneous connectivities. Likewise, Bundle Protocol (BP) of DTN architecture provides modular structure to work on top of underlying network-specific protocols through convergence layer. The flexibility of CCN and BP, however, in fact, allows us to exploit the benefits of each architecture and combine the related ideas. Integrating BP in the native CCN enhances the connectivity options of the strategy layer. Additionally, CCN can deal with network disruptions through BP.

#### C. CCNDTN protocol stack

CCNDTN protocol stack combines CCN and DTN solutions as shown in Figure 1. In this stack, DTN complements CCN forwarding mechanism from fixed networks to fragmented networks. Nevertheless, CCN explicitly support multiple connectivities (e.g., ethernet, 3G, Bluetooth, 802.11) through FIB (Figure 2). CCN is designed so that each CCN FIB entry points to a program module specialized to forwarding choices, that indicates how to forward Interests. As such,

CCNDTN includes bundle layer which enhances the strategy layer to perform in diverse environments. In fact, the strategy layer dynamically chooses interfaces from FIB entries under the changing condition. Therefore, CCNDTN router creates a FIB entry pointing to bundle daemon, once it receives prefix announcement from DTN. Subsequently, bundle layer provides seamless communication by masking the potential discontinuity, and high long delays of heterogeneous networks. This is achieved through asynchronous communication along with the use of underlying Convergence Layer Adapter (CLA).

However, BP is host centric protocol and forwarding is performed based on end-point identifier (EID). Bundle Protocol Query (BPQ) extension block has been introduced to allow an application to query a content object, which is answered by an intermediate node on the path. Intermediate nodes do not need to be addressed by destination EID. Our protocol stack uses BP and BPQ exploiting store, carry and forward mechanisms to overcome disruptions. BPQ block consists of the following fields:

- A *BPQ kind* field which indicates the bundle type. BPQ supports four types of BPQ blocks: QUERY, RESPONSE, RESPONSE_DO_NOT_FRAGMENT, PUBLISH. The third type is used in case the complete bundle is not found.
- A *BPQ value* field which contains the name of the requested content if BPQ type is QUERY or the name of the content in the payload if BPQ type is response (both type). Published bundles are treated as like response.
- An original bundle creation timestamp.
- Number of returned fragments.

As like CCN, when a BPQ enabled node receives a BPQ, the node searches its cache for BPQ responses. Subsequently, a copy of the response bundle, if available, is replied back to the node originating request. The BPQ query is not further forwarded if a complete response bundle is found. Otherwise, the node continues the search to get other fragments.

*D. Implementation*

We use CCNx [15] and DTN2 reference implementation. DTN2 is written in C++. CCNx is a reference implementation of basic CCN protocol. CCNx is written in C and Java. It also provides Android implementation for smartphones. All CCNx communication requires C implementation. We use C implementation for our prototype. We use the latest version of DTN2 2.9.0. Current version supports TCP, UDP, Nack-Oriented Reliable Multicast (NORM) and Bluetooth convergence layers (CL). It also supports external routing using external convergence layer adapter (ECLA) via XML messaging and Bundle Security Protocol (BSP) [14].

We implemented *ccn_get_dtn()* as an extension to CCN router that includes BP and BPQ. CCNx Strategy layer uses this function to transfer a request to the bundle layer. For such, strategy layer opens a connection with dtnd daemon using *dtn_open()* and registers with the daemon by *dtn_register()*. Upon successful connection and registration, strategy layer builds a bundle with a BPQ extension block. BPQ type is set as BPQ_BLOC K_KIND_QUERY and the payload contains CCN Interest packet. If any node along the path satisfies the interest, a response bundle containing the content as bundle payload and a BPQ extension describing the interest and matching content will be created and returned to the CCNDTN node that originated the BPQ. Then the strategy layer of originator performs just like native CCN router.

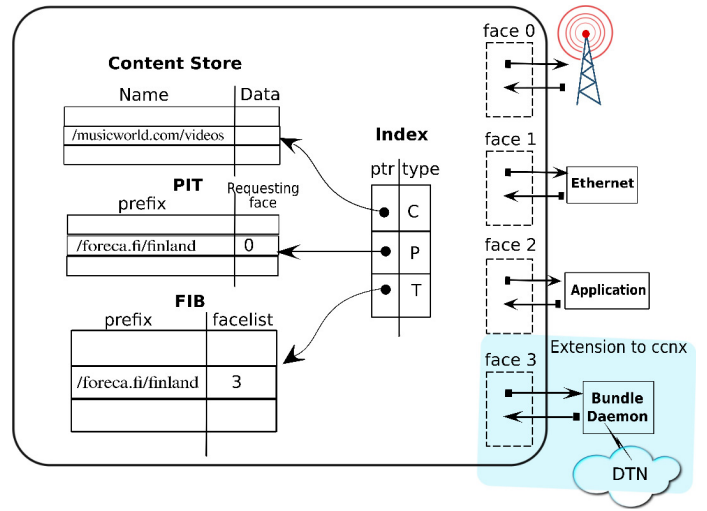

Fig. 2: CCNDTN forwarding engine.

CCN applies CCNx Repository Protocol (CCNR) [15] that provides a method for an entity such as an application to store Content Objects. A request to content storage is represented as a CCNx Interest name that specifies a prefix. If it has matching content, CCNR returns the content from its backing store. The prefix is the only required component of the Interest to look up the storage. However, we need to set the 'CCNR _DIRECTORY' environment variable with a fully qualified pathname of the repository file directory. Similarly, we need to specify repository for dtnd daemon.

IV. OPERATIONS

In this section, we briefly describe CCNDTN operations, i.e., how content distribution and retrieval works in a hybrid network. In particular, we explain the scenarios in which the switching between CCN and DTN network takes place during content distribution and retrieval.

*Content Distribution*

Let us consider a simple scenario as shown in Figure 3a. This scenario consists of both fixed network and highly dynamic network. CCN routers are fixed access points whereas CCNDTN routers are mobile nodes except E. E is an edge router.

Let us assume that a content provider next to G in DTN environment wants to publish a content. At that moment, G does not find any link to propagate announcement. The strategy layer then creates a BPQ block with type PUBLISH and transfer it to bundle layer. The bundle layer stores it in bundle cache and forwards the BPQ bundle if it gets an opportunity to transfer the announcement. Incoming announcement at any CCNDTN router is checked into local cache if there is any. Otherwise, the node creates a BPQ entry and waits for next opportunity. If any node from DTN moves to CCN network, next contact might be possible either with E or any other CCNDTN capable node. Once the prefix announcement arrived at a stable network, the strategy layer then follows standard CCN mechanisms (e.g., Intra-domain routing protocol, OSPF). On the other hand, if a node in CCN wants to publish content, the content can be distributed in CCN as well as in DTN.

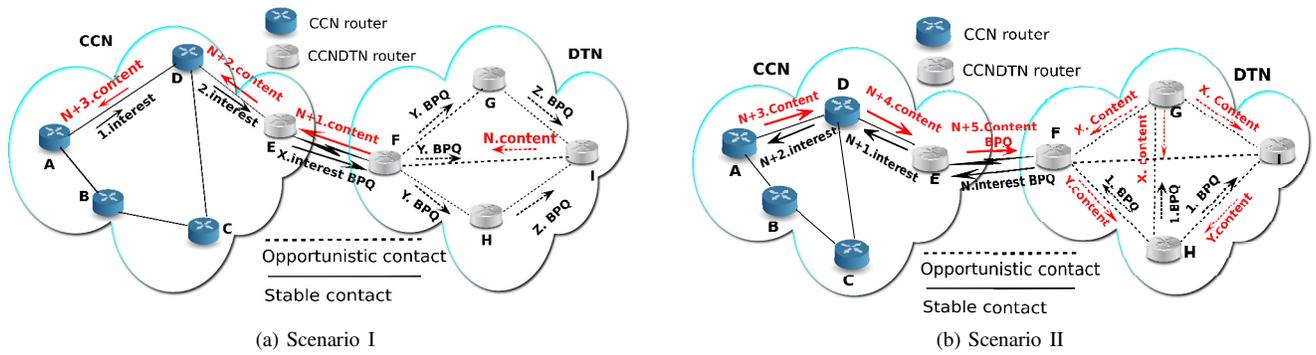

(a) Scenario I

(b) Scenario II

Fig. 3: CCNDTN operations in a simple scenario, (a) client residing in a stable network asks for a content which is in DTN environment, (b) client in a DTN environment ask for a content from a stable environment.

*Content Retrieval*

An application next to A asks for a content that resides in DTN (Figure 3b. A broadcasts the interest to all available interfaces. Subsequently, the interest packet arrives at node E (Step 1 and 2). If E chooses bundle face to forward interest, it creates a BPQ query and set BPQ value as the content name. Thus, the bundle layer take the responsibility to forward the request towards the content source. Subsequently, if available, a content source having content matching with BPQ query sends response bundle to E. Once E receives content (step 5), it replies the content back to requester (step 6 and 7) using PIT entries. In Figure 3a, X, Y, Z indicates variable number of steps (0<X,Y,Z) required for forwarding the BPQ.

Figure 3b shows the scenario where the requester is in DTN and the resource is in CCN network. For instance, any application next to H requests for a content that is located in A. Therefore, H tries to find all available link to broadcast the request. If H does not find any interface to forward the interest, it then creates a BPQ query and forward it to bundle daemon. Subsequently, the BPQ query reaches to F and transfer the request to E. Once E receives the request, it creates interest packet from BPQ query and perform standard CCN operations.As soon as, E receives the content from A, it will forward the content to originating requester H. However, if the requested content is located in DTN (e.g., in G), H receives the content after X steps (1<X).

*Experiment*

We run our prototype in six virtual machines in ubuntu 12.04 and emulate the network disruptions as shown in Figure 4. Each node runs one CCN daemon and one bundle daemon. The strategy layer of CCN can sense the connectivity and dynamically choose the available interface to forward data. In Figure 4 the dotted line indicates opportunistic link between the nodes. We intermittently start and stop the bundle daemon for realizing network disruptions. The publisher in Figure 4 announces prefix using BPQ publish. Bundle daemon in B starts. Then a link is created between publisher and B. Upon successful link creation, B hears the announcement. Now, Machine E asks for a content with BPQ query. Bundle daemon of D starts and creates a link between D and E. Daemon running in E forwards the request to D. All the opportunistic links are created using configuration file for each daemon. When A starts the daemon, E gets the content from A via D. All the daemons along the path also store the content for future requests.

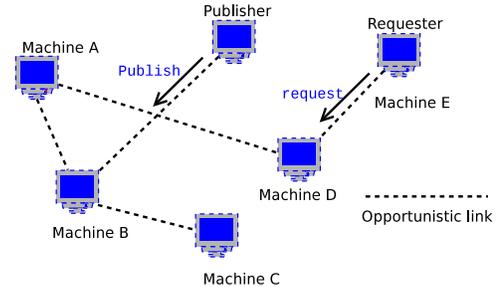

Fig. 4: Environment Setup.

## V. INTEROPERABILITY

In this section, we present the key observations towards the interoperability between CCN and DTN architecture. The summary is shown in Table II. We also discuss the conceptual overview of possible enhancements for future work.

*Name Conversion*

Name conversion is a vital question towards the interoperability between CCN and DTN. Therefore, we first need to briefly introduce the usage CCN and DTN names. CCN provides two types of packets, Interest and Data. Both packets contain a *Name* element. A name element represents a hierarchical name for CCNx content. The name element contains a sequence of *Component* elements. And each component element contains a sequence of zero or more bytes. Likewise, DTN name is expressed syntactically as Uniform Resource Identifier (URI) to identify a DTN node. Every node implementing bundle layer should have at least one EID that uniquely identifies it. Fundamentally, CCNx names identify the contents, whereas DTN names identify the DTN nodes. In conclusion, naming scheme in DTN and CCN are orthogonal. However, CCNx data formats are defined by XML schemas and use *binary encoding of XML structure*, referred as *ccnb*. In contrast, bundle protocol uses *SDNV* encoding scheme. Therefore, when CCN strategy layer sends packet using BP, it converts the CCNx packet into BPQ using SDNV encoding.

| Concern | Details | Solutions |
|---|---|---|
| Naming | Naming in DTN identifies endpoints, whereas CCN identifies contents | Orthogonal |
| Timeouts | Transition from stable to disruptive network and vice versa | Use of CCNx StatusResponse |
| Repository | CCNDTN uses two separate persistent storage for the native CCN and DTN | Abstract Content Provider on top of CCN and DTN repository (Figure 5) |
| Destination EID | Destination EID is required in DTN, whereas CCN has no notion of endpoint Id | Pseudo Destination EID with limited hop count |
| Security | CCN secures the content rather communication channel, whereas DTN uses Bundle Security Protocol | Orthogonal and complementary |

TABLE II: Questions towards CCNDTN.

*Timeouts in CCN and DTN*

Another prominent research question is the interoperability of *timeout* between CCN and DTN. CCNx includes *InterestLifeTime* into interest packet that indicates the approximate time remaining before the interest times out. Likewise, DTN architecture incorporates bundle expiration *timestamp* in each bundle. However, the question regarding timeout will arise in two cases. First, when the strategy layer of a CCNDTN router transfers the request to DTN world, it should not expect the response within a very short time. Second, when the strategy layer receives requests from bundle layer. In first case, in fact, if the final consumer does not get response within a reasonable amount of time, it retransmits the interest packet again. Consequently, unnecessary and repeated retransmissions can leave a large number of interest packets in the network. One possibility to overcome this situation is to use CCNx *StatusResponse* [15]. The StatusResponse is used to indicate an exceptional condition or additional information in response to a request. CCNx strategy layer should indicate the circumstances (e.g., status code 450: temporarily unable to complete operation) under which a StatusResponse may be returned instead of a normal response. However, in latter case, DTN timestamp could be reasonably higher than CCNx timeout. Larger timestamp could mitigate unnecessary and repeated retransmissions, since the edge router in DTN may receive the content before bundle expiration.

*Pseudo Destination EID*

In general, CCN transport API does not include any notion of source and destination addresses. In contrast, DTN2 *send and receive APIs* need Source and Destination EIDs. In fact, it is very difficult to exclude end-point EIDs from DTN architecture. For such, CCN strategy layer needs to define Pseudo destination EID which is arbitrary. Nevertheless, source EID is not difficult to define, since a node originating BPQ request can easily set its own source EID. This source EID will be used as destination EID in BPQ response. Pseudo destination EID, however, in fact, poses an inevitable constraint regarding how long or how many hops the bundle will be forwarded. Limit to hop counts may be one solution. We conclude that destination EID raises a crucial challenge towards the interoperability between ICN and DTN architecture.

*Repository Management*

CCNx and DTN2 implementation uses two separate repositories. Instead, we can design an optimized repository management to best utilize both caches and reduce memory space requirement. One possibility is designing an extra layer on top of CCN and DTN repositories as shown in Figure 5a, named as *Abstract Content Provider (ACR)*. Contents are stored in bundle cache and create references to those from CCN cache. However, this approach may lead to wrong references in CCN cache for those contents which are deleted or expired in bundle cache. For this, the expiry time of a bundle in bundle cache and reference pointers to these bundles in CCN cache must be equal. In contrast, another possibility is shown in Figure 5b. In this approach, both CCN and DTN daemon store data into a common repository. ACR of CCN and DTN performs appropriate actions (e.g., read, write) to manage the attributes of CCN packet and bundle format.

*Security*

CCN applies *content-based security*, i.e., it secures the content rather the communication channel. All contents are authenticated with digital signatures, and private contents are protected with encryption [1]. Bundle Security protocol (BSP) is specified in [14] to secure bundle. BSP uses four types of security blocks: Bundle Authentication Block (BAB), the Payload Integrity Block (PIB), the Payload Confidentiality Block (PCB), and the Extension Security Block (ESB). The BAB is used to ensure the authenticity and integrity of the bundle along a single hop from forwarder to intermediate receiver. The PIB is used to ensure the authenticity and integrity of the payload from the PIB security-source, which creates the PIB, to the PIB security-destination, which verifies the PIB authenticator. Since a BAB protects a bundle on a "hop-by-hop" basis and other security blocks may be protecting over several hops or end-to-end, BAB can be applied to our CCNDTN architecture and CCN security secures the payload. To secure BPQ extension block, ESB can be used. Therefore, BSP and CCN security is orthogonal and complements each other in DTN environment.

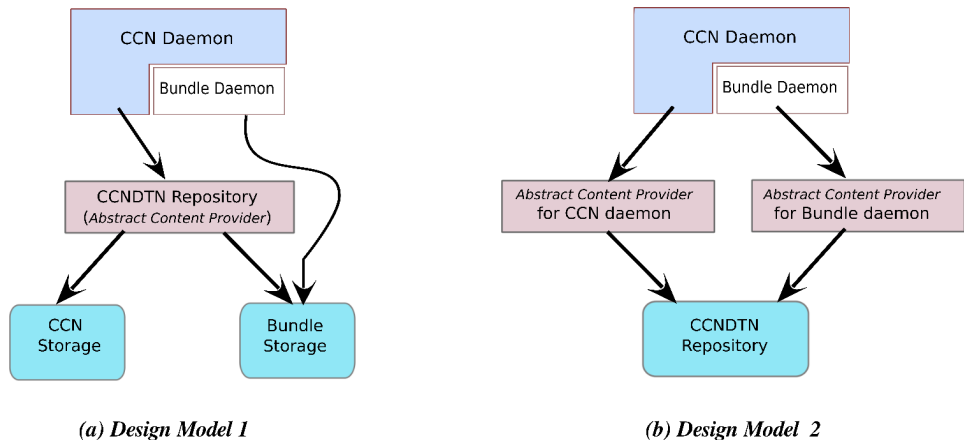

*(a) Design Model 1*  *(b) Design Model 2*

Fig. 5: Repository management for CCNDTN. (a) Abstract content provider on top of both CCN and DTN repositories, (b) Separate Abstract content providers on top of one common repository.

## VI. USE CASES

**Video on demand with DTN offloading:** CCNDTN can be effective to distribute large volume of data particularly when vehicles (e.g., high speed train, bus) can act as communication infrastructure [16]. In essence, CCNDTN can offload video on demand to alleviate the core network load. Even though mobile networks can be more effective through CCN [17], CCNDTN provides new types of service through opportunistic networking. For instance, an user can receive contents over simultaneous use of mobile network infrastructure and DTN. The strategy layer of CCNDTN can sense the connectivity state and dynamically choose available interfaces. Even if an user that was originally connected 3G cellular network moves away from 3G coverage, the user can switch to opportunistic networking of DTN. In [18], the usage of DTN has been explored in different types of network settings (e.g., Rural or developing regions, Urban areas) in which CCNDTN can be more beneficial.

**Disaster Tolerance:** In many ways we experience network disruptions in our real life. For instance, mobile networks infrastructure may not be available after a large-scale disaster (e.g., hurricane, tsunami). In such situation, CCNDTN can be very useful to disseminate disaster information through DTN.

**Mobile Sensing Application:** Now-a-days smartphones are not only used as mobile phones but also provides rich sensing capabilities through embedded sensors, such as GPS, camera, accelerometer, microphone, and digital compass. Mobile sensing applications consume a lot of power and hence, significantly have an impact on battery life. Such applications can utilize social network metrics for sharing and distributing content in a group of friends moving together in bus, train, or commuter. As for example, Haggle [19] project builds a architecture based on DTN where people with their devices utilize opportunistic contacts to exchange messages with nearby devices. However, DTN supports one connection at a time through convergence layer. In contrast, even in well connected urban area with ubiquitous availability of 3G coverage, CCNDTN will be effective through the simultaneous use of opportunistic networking of DTN and cellular networks to consume and distribute larger data items.

**Social Application:** DTN network topology exploits human connectivities in social network (e.g., PSN). PSN includes human connectivities along with physical and logical topologies in communication paradigm. Therefore, several social network metrics such as the community structure, social distance, and centrality distribution can be utilized in forwarding to experience better performance in compared to conventional Internet. Analysing social network characteristics and human connectivities, the areas of opportunistic communications can be found. In such cases, CCNDTN is advantageous through the use of DTN as well as other connectivities if available.

## VII. RELATED WORK

In the last few years, several research efforts have explored the aptness of ICN architecture with real time data (e.g., audio/video) [20] and viability in dynamic environments [17]. Listen First, Broadcast Later (LFBL) [21] has been proposed for wireless ad-hoc networks in line with named data that does not require predetermined routes, IP addressing, or a unicast-capable MAC layer. LFBL forwarding still based on implicit assumption of eventual end-to-end connectivity for a while.

In [22], the paper has explored the possibility of integrating ICN and DTN principles into a shared ICDTN architecture. The authors presented a generalized ICDTN model avoiding specific details of a particular system and provided a quantitative analysis towards ICDTN especially in PSNs. The analysis shows that ICDTN model has notable benefits in compared to DTN. In contrast, this paper focuses on the potential of integrating reference implementation of DTN architecture into CCN architecture and presents the key observations that need to be considered towards disruptions tolerant CCN.

Another key example is ICON (Information Centric Network Architecture for Opportunistic Network) [23] that incorporates some of the CCN concepts into their design. The design intent of ICON is to provide a platform through which a node in a dynamic network can be interoperable with CCN node. In contrast, our design intent is to integrate the Bundle protocol of DTN architecture in CCN node so that the strategy layer of CCN can exploit the store-carry-forward model of Bundle protocol in dynamic scenarios. NetInf project has already started to combine the related ideas of ICN and DTN architecture. NetInf will operate in DTN environment through DTN convergence layer (CL) that implements bundle protocol. NetInf did some modifications to DTN2 reference implementation for NetInf CL.

## VIII. CONCLUSION

In this paper we propose a protocol stack integrating CCN and DTN architecture to deal with network disruptions (e.g., large RTT, network partitioning). It is important to note that high delay tolerance is not necessarily weakness in CCN. Instead, by integrating Bundle Protocol of DTN architecture in the native CCN will enrich the connectivity options of CCN architecture in fragmented networks. Furthermore, this paper highlights the potential use cases, where CCNDTN can be more beneficial through the simultaneous use of cellular networks and opportunistic networking for the dissemination of larger data items. While discussing the implementation details and the feasibility of the prototype, we address the crucial questions about integrating Bundle protocol in CCN and propose possible enhancements for Future work.